\documentclass[11pt,twoside]{article}

\usepackage{graphicx}
\usepackage{asp2006}
\usepackage{epsf}
\usepackage{lscape}

\begin{document}
\title{Spectroscopic Mode Identification in Slowly Pulsating Subdwarf-B Stars}   
\author{Caroline Schoenaers and Tony Lynas-Gray}   
\affil{Department of Physics, University of Oxford, Denys Wilkinson Building, Keble Road, Oxford OX1~3RH, United Kingdom}    

\begin{abstract} 
Mode identification is crucial for an asteroseismological study of any significance. Contrarily to spectroscopic techniques, methods such as period-fitting and multi-colour photometry do not provide a full reconstruction of non-radial pulsations. We present a new method of spectroscopic mode identification and test it on time-series of synthetic spectra appropriate for pulsating subdwarf-B stars. We then apply it to the newly discovered slowly pulsating subdwarf-B star HD~4539.
\end{abstract}

\section{Introduction}
Subdwarf B (sdB) stars are low-mass $(\sim 0.5 M_{\odot})$ core helium 
burning objects with thin and mostly 
inert hydrogen-rich residual envelopes. They belong to the Extreme Horizontal Branch (EHB) \citep{heber1986} and remain hot 
$(20000 \le T_{\rm eff} \le 40000)$ and compact $(5 \le {\log g} \le 7)$ throughout 
their lifetime \citep{saffer1994}, before evolving towards the white dwarf cooling sequence without 
experiencing the Asymptotic Giant Branch and Planetary Nebula phases of stellar evolution. 
\citet{han2002,han2003} use binary population synthesis calculations to demonstrate the formation of sdB stars through several possible channels, but resulting models require further comparison with observation. \citet{kilkenny1997} and \citet{green2003} respectively discovered that some sdB stars undergo fast and slow non-radial pulsations. This means asteroseismology can be used to study the internal structure of these stars and so constrain evolution models. 

After having measuring the pulsation periods, the first step of any seismic modelling is to identify the pulsation modes that are excited\footnote{Since non-radial pulsations are modelled by spherical harmonics $Y_{n\ell m}$, mode identification consists in assigning values to the spherical wavenumbers $n$, $\ell$ and $m$, respectively the number of nodes of the radial displacement, the number of nodal lines on the stellar surface and the number of such lines passing through the rotation axis of the star.}, which is far from being trivial. In the case of pulsating sdB stars, most mode identifications up to date have been performed using either period fitting \citep[\emph{etc.}]{brassard2001,charpinet2005,randall2006a,randall2006b} or multicolour amplitude ratios \citep[\emph{etc.}]{jeffery2004,jeffery2006}. However, these methods do not provide a full reconstruction of the pulsations, and should therefore be complemented by spectroscopic mode identification techniques, such as the one we present here.
  
\section{Spectroscopic Mode Identification}
As reported earlier \citep{schoenaers2006}, line-profile variations (lpv), although weak, are to be expected in pulsating subdwarf-B stars. Furthermore, they have been observed, for instance in PG~1325+101 by \citet{telting2004}. A reliable spectroscopic mode identification method is therefore needed to take advantage of these and similar observations.
\subsection{Why not use the moment method?}
Nowadays, the most commonly used spectroscopic mode identification method is the ``moment method'' (hereafter MM). It was first introduced by \citet{balona1986a,balona1986b,balona1987} and later reformulated into computer-friendly terms and applied to Main Sequence pulsators by \citet{aerts1992} and \citet{aerts1996} for monoperiodic stars, while \citet{briquet2003} extended the formalism to multiperiodic stars. In this method, one does not try to identify pulsation modes directly from the line-profiles, because this is often impossible for multiperiodic stars, but instead replaces each observed line-profile by its first few moments and compares their time dependence with that of theoretical moments. The latter depending on the wavenumbers $\ell$ and $m$ of the pulsation mode, as well as on the pulsation amplitude and on the inclination angle $i$ of the star, this method should in principle allow a complete reconstruction of any non-radial pulsation. It has been used successfully for the study of $\beta$ Cephei and Slowly Pulsating B (SPB) stars \citep{briquet2003,decat2005}. However, this method makes use of a crucial (and not always justified) approximation: except for radial velocity variations, the MM neglects all other variations due to non-radial pulsations. 

However, because of non-radial pulsations, different points on the stellar surface not only have different radial velocities, but also different temperatures, $\log g$ and orientations, and their contributions to a line-profile therefore have different amplitudes. Hence, lpv do not only come from the Doppler shift in wavelength, but also from the complex temperature and $\log g$ behavior on the stellar surface. This is especially the case in sdB pulsators: because they are so dense, any significant displacement of the stellar surface due to pulsation is kept at its minimum, and instead the available energy causes temperature perturbations in the stellar photosphere. To demonstrate that neglecting temperature, gravity and geometric variations can lead to a mistaken mode identification, even in the simplest cases, we used the MM to identify various monoperiodic pulsations present in time-series of synthetic spectra computed following \citet{schoenaers2006}. 
On the one hand, it could be seen that the MM always provide, within its five\footnote{Following \citet{briquet2003}'s advice.} best solutions, the proper mode identification (as well as very good estimates of $i$, $A_\mathrm{p}$ and $v_{\rm eq}$) when only radial velocity variations were taken into account. However, on the other hand, when temperature, gravity and geometric variations were taken into account (as indeed they should be) this test demonstrated that even in the simple case of monoperiodic pulsations, the current MM falls short of properly identifying all pulsational characteristics. Clearly, in the case of pulsating sdB stars, where temperature, pressure and geometric effects have to be taken into account, an improved version of the moment method is needed.

\subsection{The ``synthetic moment'' method}
\citet{schoenaers2006} introduced a state-of-the-art code that computes times-series of synthetic spectra and lpv. The ``synthetic moment'' method (hereafter SMM) introduced here takes advantage of this ability to compute ``synthetic moments'' for these extremely accurate synthetic lpv, and to compare them to observed moments: for a given target and observed line-profile variation, a grid of time-series of synthetic spectra spanning a suitable parameter space is computed, the first three moments of each of these synthetic lpv are obtained, and the best fit between observed and synthetic moments is found by minimising a merit function of the form
\begin{eqnarray}
\Delta_{\ell',m'} &=& \Bigg\{ \frac{1}{N_\mathrm{obs}} \sum_{j=1}^{N_\mathrm{obs}} \Big[ \big(<v>_{\mathrm{syn},j}-<v>_{\mathrm{obs},j}\big)^2 \nonumber\\
&\phantom{=}&\phantom{12345678} +\big|<v^2>_{\mathrm{syn},j}-<v^2>_{\mathrm{obs},j} \big| \\
&\phantom{=}&\phantom{12345678} +\big(<v^3>_{\mathrm{syn},j}-<v^3>_{\mathrm{obs},j}\big)^{2/3} \Big]\Bigg\}^{1/2}\nonumber
\end{eqnarray}
where $N_\mathrm{obs}$ is the number of observations, odd moments have average zero, and the synthetic second moment $<v^2>_{\mathrm{syn},i}$ is adjusted to have the same average as its observed counterpart $<v^2>_{\mathrm{obs},i}$.

The choice of the parameter space to be searched is crucial: if too limited one might miss the proper mode identification, but if too broad the computing time of the synthetic lpv can become prohibitive. One could consider building an extensive grid of synthetic lpv by varying these parameters more or less ``continuously'', but because most parameters are target-dependent, it is simply more efficient to compute one grid for each pulsating star. Provided spectroscopic observations of the target are obtained at high enough SNR, $T_{\rm eff}$, $\log g$, $v_{\rm eq}\sin i$ and $[\mathrm{X}/\mathrm{H}]$ can be determined to good accuracy. Before attempting mode identification, one should also obtain reliable values of the pulsation period(s) $P$, either through (ideally simultaneous) photometry or radial velocity measurements. If the latter are obtained (either directly through spectroscopic measurements or indirectly from the first moment), then good estimates of the physical velocity amplitudes of the pulsation mode(s) can be obtained; if not, then pulsation amplitudes can in principle remain a free parameter, along with the inclination $i'$ and the actual mode identification $(\ell',m')$.

Ideally, when dealing with a multiperiodic star, all pulsation modes should be identified simultaneously, but this would require a very large grid of synthetic lpv, which can become prohibitive. It is however possible to attempt mode identification as follows:
\begin{enumerate}
\item Use the SMM on the dominant pulsation mode with a suitable primary grid of synthetic lpv to obtain $(\ell'_1,m'_1)$ and $i'$ for the five best fitting solutions (minimising $\Delta_{\ell'_1,m'_1}$). Then, if at all possible, further discriminate between these five best solutions (see next subsection for more details).
\item For each of the possible $i'$ and $(\ell'_1,m'_1)$ combinations obtained in Step 1, build a secondary grid of synthetic lpv (one for each $(\ell'_2,m'_2)$ combination) and use the SMM to identify the second pulsation mode.
\item Step 2 can be repeated at will until all modes have been succesfully identified.
\end{enumerate}

\subsection{Discriminating between the few best solutions}
As experience has shown in the case of Main Sequence pulsators (see for example \citet{decat2005}) it is in general safer not to accept straight away the best solution of the MM (and by extension, of the SMM) as the proper mode identification, but instead to discriminate between a few best solutions using some other mode identification method. Similarly, to confirm the identification of pulsation modes in PG1716 stars, we suggest a visual inspection of the fit between observed and synthetic moments, as well as the use of the IPS (for Intensity Period Search) diagnostic \citep{schrijvers1997,telting1997a,telting1997b,schrijvers1999} to compare the amplitude and phase diagrams of observed lpv with those of synthetic profiles corresponding to the few best solutions of the SMM.

\subsection{Test on synthetic data}
\begin{figure}[!ht]
\centering
\begin{minipage}{13cm}
\begin{minipage}{6.4cm}
\centering
\includegraphics[scale=0.25,clip]{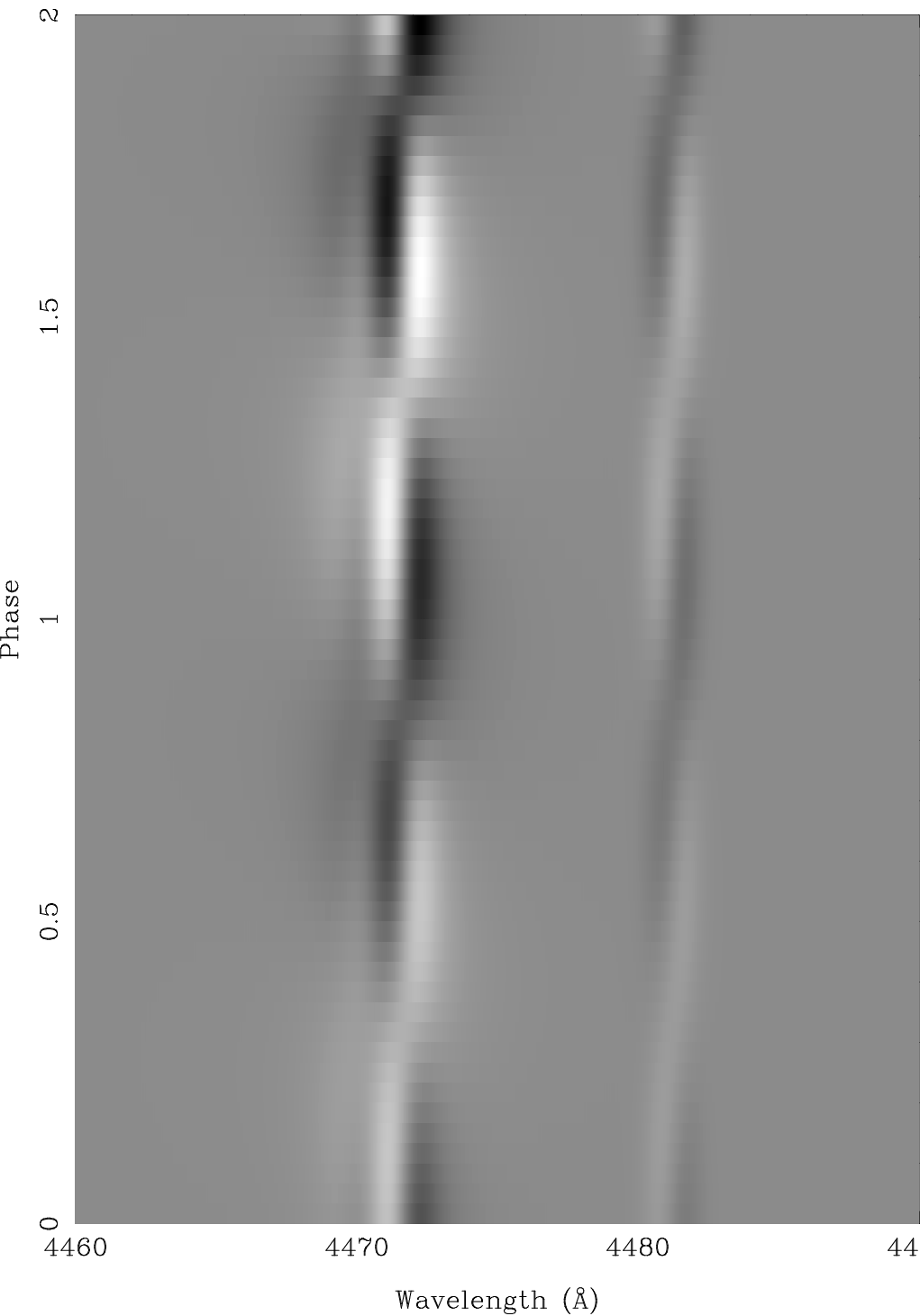}
\end{minipage}
\begin{minipage}{6.4cm}
\centering
\includegraphics[scale=0.25,clip]{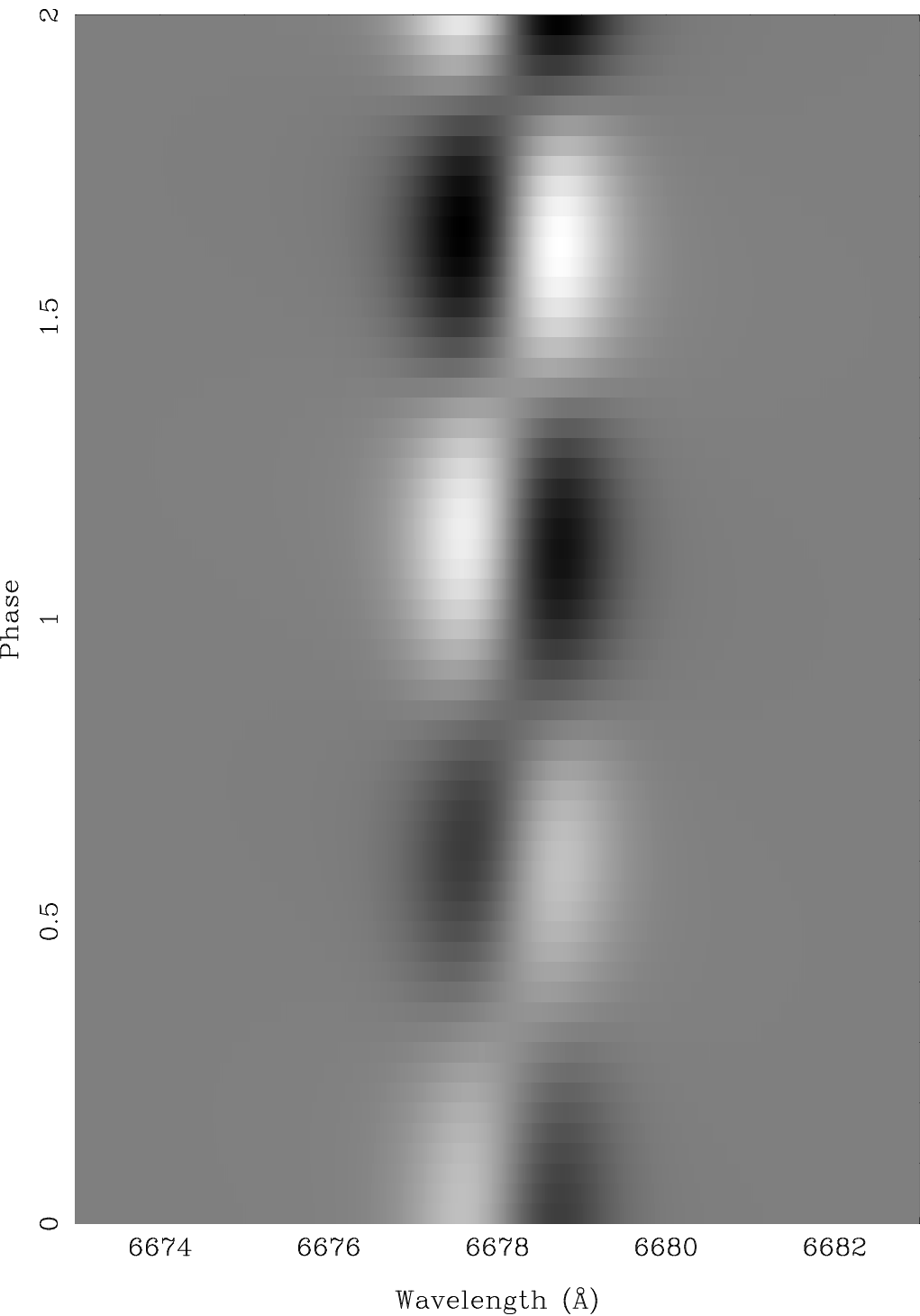}
\end{minipage}
\end{minipage}
\caption{Grey-scale plots of synthetic lpv (\emph{left}: He~I $\lambda 4471\,$\AA\ and Mg~II $\lambda 4481\,$\AA, \emph{right}: He~I $\lambda 6678\,$\AA) with periods and $\ell$-values representative of PG~1338+481 \citep{randall2006a}: $i=60\deg$, $(\ell_1,m_1)=(1,1)$, $P_1=3828\,$s, $A_{\rm p,1}=2.0\,{\rm km}\cdot{\rm s}^{-1}$ and $(\ell_2,m_2)=(1,0)$, $P_2=3530\,$s, $A_{\rm p,2}=1.4\,{\rm km}\cdot{\rm s}^{-1}$.}\label{synthlpv}
\end{figure}

Figure \ref{synthlpv} shows in grey-scale a biperiodic lpv, representative of the PG1716 star PG~1338+481 ($T_{\rm eff}=27\,000$\,K, $\log g=5.4$, $M=0.48\,M_{\sun}$, $R=0.23\,R_{\sun}$, $v_{\rm eq}=5\,{\rm km}\cdot{\rm s}^{-1}$, [He/H]$=-1.3$, [Mg/H]$=-1$), that was used (amongst many others) to test the SMM. Pulsation characteristics are given in the caption. 
Appropriate non-adiabatic parameters $R$ and $\psi_T$ were computed using \citet{randall2005}'s Equations 46 and 47\footnote{Note that Equation 47 of \citet{randall2005} should read $\langle\psi_T\rangle=\pi-2.30\exp(-P\ell^{0.38}/1800)$.}

\begin{figure}[!ht]
\centering
\includegraphics[clip,scale=0.5]{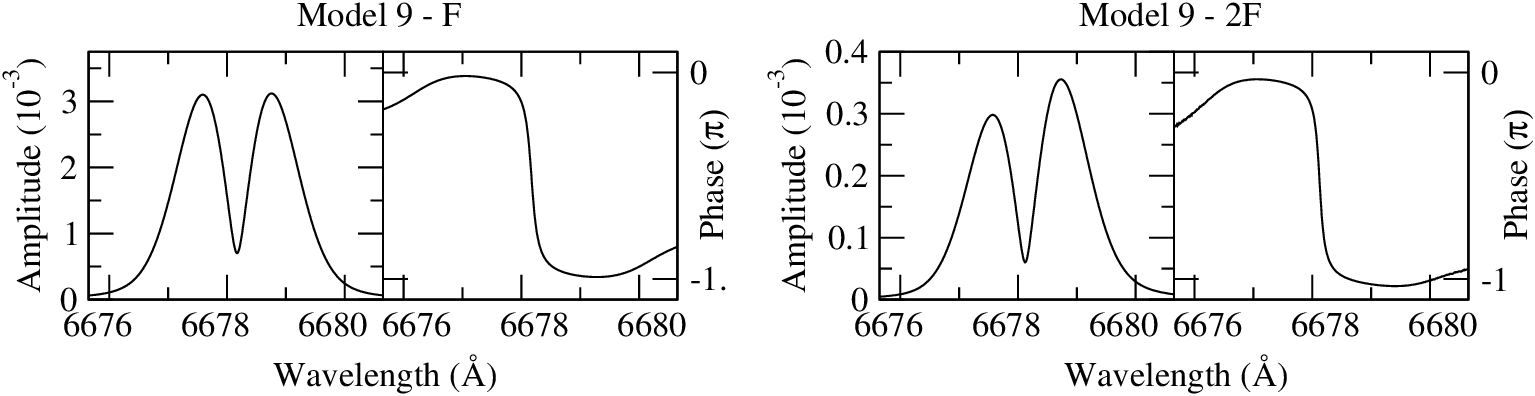}
\caption{``Observed'' IPS diagnostic for the lpv shown in Figure \ref{synthlpv}. The two leftmost panels show the amplitude (\emph{left}) and phase (\emph{right}) of variations with the main frequency across the profile, while in the two rightmost panels the amplitude (\emph{left}) and phase (\emph{right}) plots are for variations with the first harmonics of the main frequency.}\label{mod9ips}
\end{figure}

\begin{figure}[!ht]
\centering
\includegraphics[clip,scale=0.5]{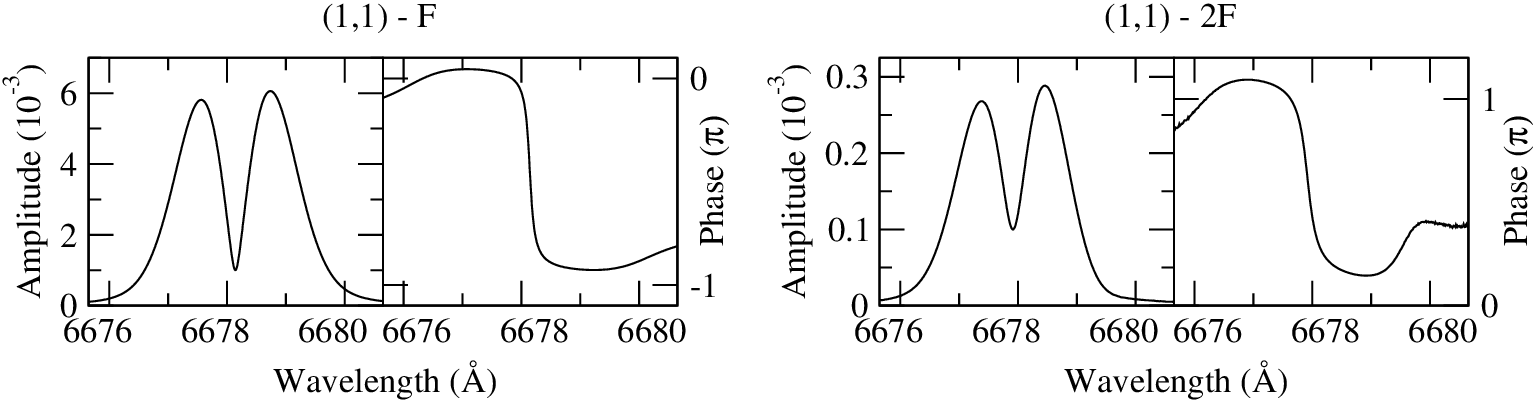}
\includegraphics[clip,scale=0.5]{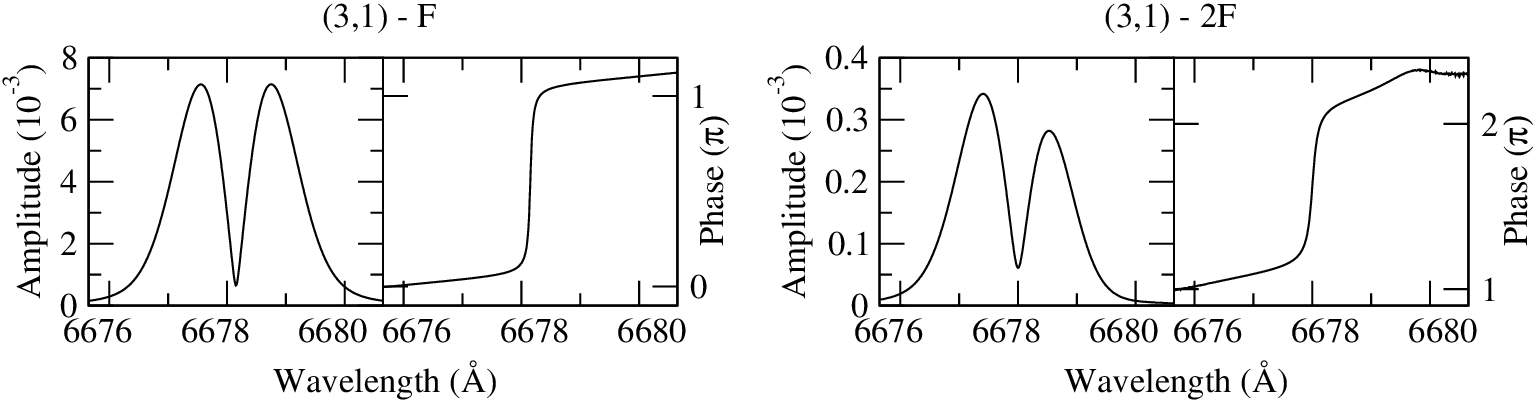}
\caption{Same as Figure \ref{mod9ips}, but for possible identifications of the first pulsation mode: $(\ell'_1,m'_1)=(1,1)$ (\emph{top}) and $(3,1)$ (\emph{bottom}).}\label{model9_1}
\end{figure}

Applying the SMM to these synthetic lpv of He~I $\lambda 6678\,$\AA\footnote{Similar results are obtained when applying the SMM to He~I $\lambda 4471\,$\AA\ and Mg~II $\lambda 4481\,$\AA\ \citep{schoenaers2008}.} provides, for the first pulsation mode, the five best solutions listed in the first four columns of Table \ref{SMM_id_bi}.
\begin{table}
\caption{Five best identifications obtained by the SMM for the biperiodic pulsation modelled above. The first column is the inclination of the rotation axis with regard to the line-of-sight (in degrees), the second and third columns give the $(\ell_1',m_1')$-values of the first mode, and the fourth provides the value of $\Delta_{\ell_1',m_1'}$, minimum for the best fitting solution. Columns 5 to 7 give the $(\ell_2',m_2')$-values and the value of $\Delta_{\ell_2',m_2'}$ for the second mode, provided the first was properly identified. For each mode, the proper identification is written in bold font.}\label{SMM_id_bi}
\smallskip
\begin{center}
\begin{tabular}{c ccc c ccc}
\tableline
\noalign{\smallskip}
&\multicolumn{3}{c}{First mode}&~~~~&\multicolumn{3}{c}{Second mode}\\
\noalign{\smallskip}
\cline{1-4}\cline{6-8}
\noalign{\smallskip}
$i'$ & $\ell'_1$ & $m'_1$ & $\Delta_{\ell'_1,m'_1}$ &&$\ell'_2$ & $m'_2$ & $\Delta_{\ell'_2,m'_2}$\\
\noalign{\smallskip}
\tableline
\noalign{\smallskip}
60 & \textbf{1} & \textbf{1} & \textbf{4.5436200} && \textbf{1} &$\phantom{-}$\textbf{0} & \textbf{1.31584586} \\
30 & 2 & 1 & 5.1651505 && 1 &$-1$ & 1.90534104 \\
30 & 3 & 1 & 5.2640002 && 3 &$\phantom{-}2$ & 2.25563231 \\
85 & 2 & 0 & 5.4004608 && 3 &$-3$ & 2.43201440 \\
60 & 3 & 0 & 5.4952073 && 3 &$-2$ & 2.44369818 \\
\noalign{\smallskip}
\tableline
\end{tabular}
\end{center}
\end{table}
Visual inspection of the match between ``observed'' and  synthetic moments (not plotted here for the sake of brevity, but see \citet{schoenaers2008}'s Figure 4.23) allowed us to rule out all but the $(\ell'_1,m'_1)=(1,1)$ and $(3,1)$ identifications for the first mode. Further discrimination can be achieved by comparing the ``observed'' IPS diagnostic (shown in Figure \ref{mod9ips}) to those of the $(\ell'_1,m'_1)=(1,1)$ and (3,1) identifications (plotted in Figure \ref{model9_1}).
The phase diagrams of (3,1) do not exhibit the right behaviour, while those of (1,1) do. The latter is therefore selected as the identification of the dominant mode.

\begin{figure}[!ht]
\centering
\includegraphics[clip,scale=0.5]{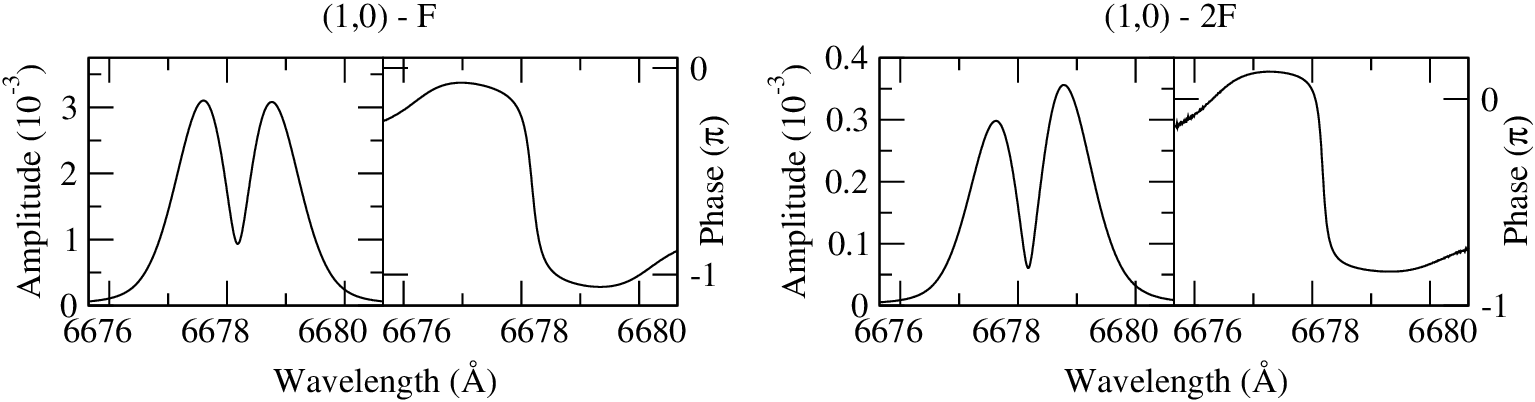}
\caption{Same as Figure \ref{mod9ips}, but for possible identifications of the first pulsation mode: $(\ell'_1,m'_1)=(1,1)$ (\emph{top}) and $(3,1)$ (\emph{bottom}).}\label{model9_2}
\end{figure}

Keeping $(\ell_1,m_1)=(1,1)$ and $i=60\deg$ fixed, and applying the SMM to identify the second pulsation mode yields the five best identifications listed in Columns 5 to 7 of Table \ref{SMM_id_bi}. The IPS diagnostic of the best identification $(\ell'_2,m'_2)=(1,0)$ plotted in Figure \ref{model9_2} matches the ``observed'' IPS diagnostic very well, 
and this shows the SMM does provide a reliable mode identification, even for multiperiodic stars.

\section{Application to a New Slowly Pulsating Subdwarf-B Star: HD~4539}
\begin{figure}
\centering
\begin{minipage}{13cm}
\begin{minipage}{6.4cm}
\centering
\includegraphics[clip,scale=0.25]{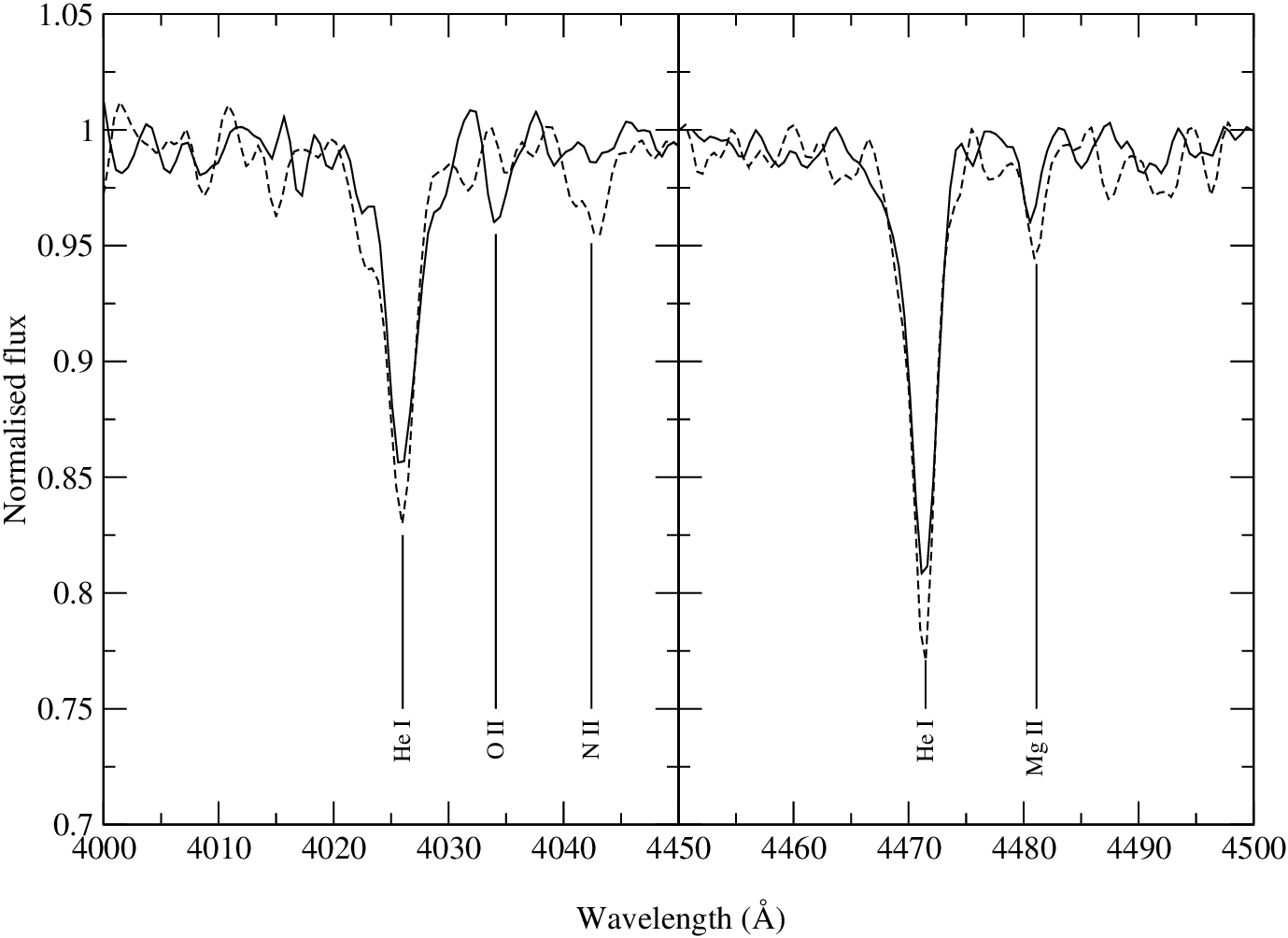}
\end{minipage}
\begin{minipage}{6.4cm}
\centering
\includegraphics[clip,scale=0.25]{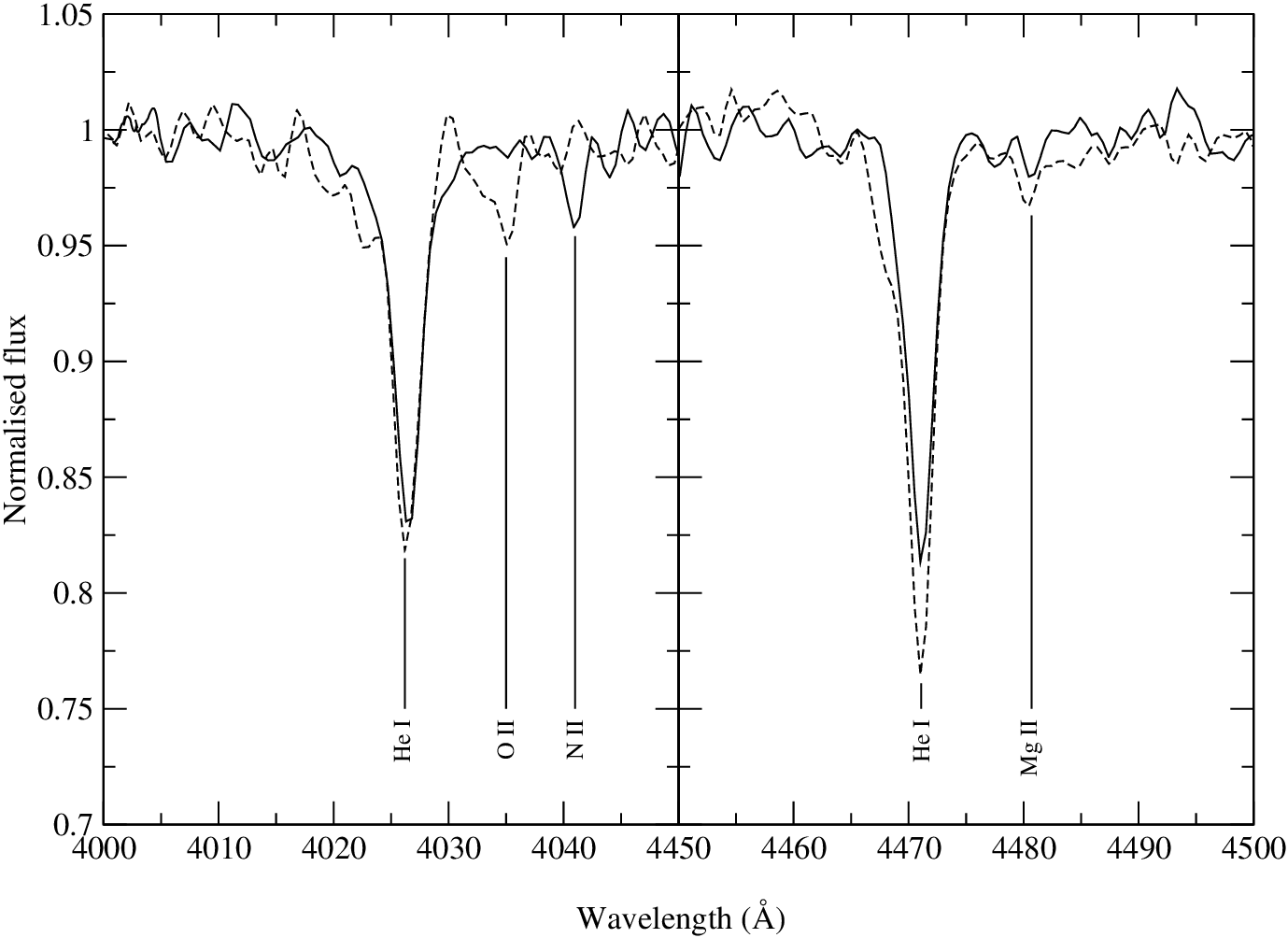}
\end{minipage}
\end{minipage}
\caption{Line-profile variations in HD~4539 observed in August 2005 (\emph{two leftmost panels}) on MJD=53611.13918 (\emph{solid line}) and MJD=53612.03206 (\emph{dashed line}), and in August 2006 (\emph{two rightmost panels}) on MJD=53958.15420 (\emph{solid line}) and MJD=53952.16671 (\emph{dashed line}).}\label{HDlpv}
\end{figure}

HD~4539 ($B\sim 10.12$, 2000.0 coordinates: $\alpha=0^\textrm{h}47^\textrm{m}29.2^\textrm{s}, \delta=+09^\circ 58' 56''$) is one of the brightest known sdB stars. Its surface parameters \citep[see e.g.][]{bascheck1972,saffer1994} are similar to those of PG1716 stars, and \citet{schoenaers2007} report the discovery of non-radial pulsations in HD~4539, together with their frequency analysis, but without mode identification and subsequent analysis. In Figure \ref{HDlpv}, we plot lpv observed in HD~4539 in August 2005 and 2006.
These lpv are much stronger than the average noise present in our spectra, indicating they are real. A careful radial velocity analysis and period search yielded four frequencies \citep{schoenaers2007}, of which we identify the dominant one ($P_1=9310\,$s) in this preliminary report. A  more thorough analysis will be presented elsewhere.

The wavelength range of our spectra includes He~I $\lambda$ 4471\,\AA~and Mg~II $\lambda$ 4481\,\AA, and we therefore applied the SMM to the first three moments of these two lines. In Table \ref{idHD} are listed the five best-fitting identifications of the dominant pulsation mode.
\begin{table}
\caption{Five best fitting SMM identifications of $P_1$ in HD~4539, using He~I $\lambda$ 4471\,\AA~(\emph{left}) and Mg~II $\lambda$ 4481\,\AA~(\emph{right}). In each case, the first and second columns give the tentative $(\ell'_1,m'_1)$ identification, the third column provides the corresponding inclination, and the fourth is the merit function $\Delta_{\ell'_1,m'_1}$.}\label{idHD}
\smallskip
\begin{center}
\begin{tabular}{cccc c cccc}
\tableline
\noalign{\smallskip}
\multicolumn{4}{c}{He~I $\lambda$ 4471\,\AA} && \multicolumn{4}{c}{Mg~II $\lambda$ 4481\,\AA}\\
$\ell'_1$ & $m'_1$ & $i'$ & $\Delta_{\ell'_1,m'_1}$ &~~~~&$\ell'_1$ & $m'_1$ & $i'$ & $\Delta_{\ell'_1,m'_1}$\\
\noalign{\smallskip}
\tableline
\noalign{\smallskip}
1 & 1 & 85 & 35.7424808 && 3 & 3 & 85 & 55.5834533 \\
2 & 1 & 60 & 35.7468680 && 3 & 1 & 30 & 55.5841337 \\
2 & 0 & 85 & 35.7529125 && 2 & 0 & 85 & 55.5845978 \\
3 & 3 & 85 & 35.8536815 && 2 & 1 & 60 & 55.5993271 \\
3 & 1 & 30 & 35.8943103 && 1 & 1 & 85 & 55.6007015 \\
\noalign{\smallskip}
\tableline
\end{tabular}
\end{center}
\end{table}
These five best solutions (including the inclination) are the same (although found in a different order) whether the identification is performed using He~I $\lambda$ 4471\,\AA~or Mg~II $\lambda$ 4481\,\AA, which already indicates they are quite robust. Visual inspection of the moments did not allow us to discard any particular identification, and therefore we compared the IPS diagnostic of all five best-fitting solutions to the observed IPS diagnostic of HD~4539, computed for both He~I $\lambda$ 4471\,\AA~and Mg~II $\lambda$ 4481\,\AA. The same conclusion could be drawn in both cases, namely that the dominant pulsation mode ($P_1=9310\,$s) in HD~4539 can be identified as a $\ell_1=1, m_1=1$ mode, while the inclination of the rotation axis with respect to the line-of-sight is (close to) $85\deg$.

\acknowledgements The authors wish to thank the editors for allowing a very late submission. C.S. thanks St Cross College, the MPLS division of the University of Oxford, and the LOC for their generous contribution to my expenses during the Third Meeting on Hot Subdwarf Stars.

\end{document}